# Convolutional Neural Network Approach for Emitter Association using Channel Identification in a MIMO System based on Propagation Features

Michael F. Talley, Jr.
Electrical and Computer
Engineering
Morgan State University
Baltimore, USA
mital4@morgan.edu

Kofi Nyarko, D.Eng
Electrical and Computer
Engineering
Morgan State University
Baltimore, USA
kofi.nyarko@morgan.edu

Willie L. Thompson, II, D.Eng.
Electrical and Computer
Engineering
Morgan State University
Baltimore, USA
willie.thompson@morgan.edu

Arlene Cole-Rhodes, Ph.D.
Electrical and Computer Engineering
Morgan State University
Baltimore, USA
arlene.colerhodes@morgan.edu

Craig Scott, Ph.D.
Electrical and Computer Engineering
Morgan State University
Baltimore, USA
craig.scott@morgan.edu

*Abstract*— **In this paper, an application of a 1D deep convolutional neural network (DCNN) and 4x4 1D DCNN Multi-channel Model (DCNN-MCM) was developed to predict the probability of a channel being associated with a given transmitter for each emitter in a 4x4 multi-input multi-output (MIMO) system. Counterintuitively, compared to the traditional approach to emitter association (EA), this research argues for the identification of received RF signals based on channel features (CFs) such as the channel impulse response (CIR) and transfer function (TF). Based on the CFs there are unique properties per transmit and receive pair that can be used to identify the different transmitters. More specifically, CFs are often defined by a wide sense stationary (WSS) stochastic process which gives them unique properties such as, being statistically independent and reciprocal for emitter association. Given these CFs, the received RF signals can be used to measure the clutter and objects surrounding the emitters. Safely assuming the surrounding clutter and objects are unique to each emitter, the estimation and tracking of this clutter provides a way of identifying what emitter is transmitting the observed signal. This method successfully classifies and identifies the emitters at 97.22% and 88.89% accuracy for the DCNN and DCNN-MCM respectively. This method was further compared with the accuracy of previous methods to test the validity of the approach and future improvements.**

## I. Introduction

Achieving information superiority is vital for military decision making. The increasingly complex electromagnetic spectrum is a key source of information for both strategic and tactical intelligence. Today's military challenges include how to effectively extract, classify, and locate transmissions within the signal spectrum. Using deep learning (DL) systems, these signals can be classified as friend or foe and by type of equipment, then geolocated to gain intelligence on who, what, and where the signal came from. Given the ever changing, complex, and crowded wireless spectrum the military will need the ability to control and dominate these environments if they wish to maintain their preeminence in achieving control in combat operations. To do this, there needs to be a continual evolution and innovation of the technique that is referred to as electronic warfare support (ES). ES is a subdivision of electronic warfare (EW) which involves actions taken by the warfighter to detect, intercept, identify, locate, and/or localize sources of intended and unintended radiated electromagnetic energy or emitters. By monitoring the utilization of emitters, the warfighter can make strategic decisions on opposing force intent.

Radar (emitter) systems measure range, angle and velocity of objects leveraging waves reflected from said object. These reflected signals can be continuous waves or pulsed. The radar system can extract characteristics from the signals, which can include amplitude, frequency, direction, angel-of-arrival, etc. With this information, a signal descriptor word (SDW) is created, which contains measured and derived signal features that can be used for emitter identification. For pulsed signals, this descriptor word is referred to as a pulse descriptor word (PDW). The reflected pulses can then be sorted into groups, called bursts, where each burst contains a PDW associated with a single emitter. A group of similar bursts represents the observed emitters, and each emitter gets a description for specific emitter identification (SEI).

The traditional methods for SEI have proven to be successful but are becoming more complex given the process that researchers must undertake to continue yielding the same performance. According to [1] there are four specific limitations that have been identified in research for SEI traditional methods listed below but this research alternative has proven that feature-engineering is not needed to yield the same results. These limitations are 1) traditional SEI approaches are the extraction and use of pre-determined and expert-defined features, 2) traditional SEI approaches only consider specific aspects of the received signal, 3) feature extraction often requires pre-processing of the received signal, including synchronization, carrier tracking, demodulation, and SNR estimation, in addition to the computational cost of extracting the expert features, and 4) unless in a cooperative environment, the number of clusters (emitters) from clustering algorithms will not be known in practice and is subject to change, severely limiting the ability to identify anomalous emitters and behaviors.

In contrast, this work aims to utilize channel features to perform channel identification for SEI and emitter association (EA) instead of the PDW. The primary focus is to explore the utilization of DL for the EA within the multiple-input multiple-output (MIMO) environment, as illustrated in Fig. 1. For this research EA is a two-step process: 1) channel identification of the transmitter and received pair and 2) the association of the channel with a given transmitter. To extract the features the cross-correlation of the MIMO channel response matrix of the environment will be investigated.

From the channel response matrix, a channel estimation dataset was collected using a 4x4 MIMO receiver system based on Zadoff-Chu (ZC) sequences. Given the transmitted RF signal, this work will analyze the data to identify the salient features for specific propagation channel characteristics. Using the data-driven characteristics of the channel impulse response (CIR) and the transfer function (TF), or the channel features (CFs) of the channel, a 1D Deep Convolutional Neural Network (DCNN) and 4x4 1D DCNN Multi-channel Model (DCNN-MCM) was developed to predict the probability of identification for each channel associated with each emitter. Using the features from this analysis in a DL algorithm will aid in the classification of the channels for each emitter of interest. This method was further compared with the accuracy of previous methods to test the validity of the approach.

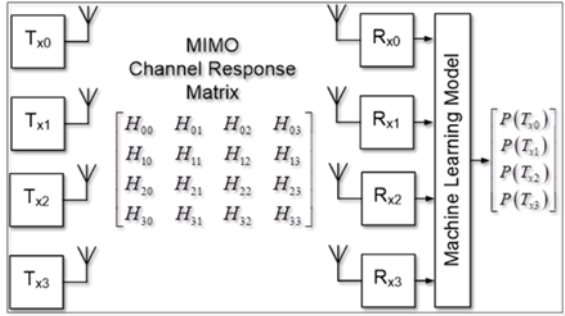

Fig. 1. MIMO Channel Response System.

### A. Related Work

Within a military ecosystem, EW represents the ability to use the electromagnetic spectrum—signals such as radio, infrared or radar—to sense, protect, and communicate. At the same time, it can be used to deny adversaries the ability to either disrupt or use these signals [2]. This ability is crucial to defending and aiding in SEI for the warfighter during mission critical efforts. During mission critical efforts it is important for the warfighter to be able to analyze the surrounding area and know what different entities are there – and to determine whether they are allies or enemies. A specific form of this can be defined as SEI. SEI is the process of analyzing electromagnetic signals to determine the emitters' location and whether it is hostile or friendly [3]. EA is a method that is used in SEI for the association of RF signals to different emitters.

There are three (3) different ways to approach this problem: the traditional way, machine learning (ML) with expert features found in the traditional way, and DL with a data driven approach. With all three approaches there is an "all" or "combination four main components" approach. These four main components consist of: preprocessing, feature generation, classification, and identification. While the traditional way has proven to be successful, the Army Research Lab (ARL) has identified gaps in the current approach that will only get more complicated as technology evolves [4]. Thus, there are advantages to the ML and the combination of the two. The advantage of the data driven DL approach is that the entire process can be made automatic and can be deployed to autonomously classify the signals within an environment.

Traditionally SEI seeks to perform the technique of EA using expertly engineered features from the PDW [2]. Within the PDW an emitter has specific signal characteristics that can be employed to identify the specific emitter and location. These signal characteristics or parameters can be divided into two groups: primary and secondary. The signals' primary characteristics are carrier frequency (RF), time of arrival (TOA), pulse width (PW), angle of arrival (AOA) and amplitude (A). The secondary characteristics, which are derived from signal processing, are: pulse repetition interval (PRI), polarization, variation of PRI (agility, jitter, stagger), variation (agility) of RF, scan type (ST), and scan period (SP) [3]. Taking into consideration the fact that the emitter signal recognition and identification process is an integral part of contemporary combat operations, the process must continue to evolve and improve.

For ML there can be two different approaches: combine the knowledge of expert features with ML, or a data driven approach. ML is an application of artificial intelligence (AI) that provides systems the ability to automatically learn and improve from experience without being explicitly programmed [5][6]. The first approach consists of a ML algorithm that takes expert features as an input - known as a feature vector. The feature vector will consist of the most salient features for the proposed task. Given the feature vector the ML algorithm can learn from it and then produce an output for classifying and identifying specific emitters. The expert features that are needed to perform this task can be found in the digital signal package known as PDW. The second approach consisting of ML is in slight contrast to the first approach. A data-driven approach is based on the analysis of the data about a specific system. The main concept of a data-driven model is to find relationships between the input (signal data/features) and the output (which emitter) without explicit knowledge of the physical behavior of the system [6].

Given these two approaches there are three specific ways and papers that are pertinent to the approach of this research. The first approach, from [7], used a CNN) to learn the features/characteristics of the signals for the task of classification and identification. Their approach used three main parameters of a radar signal for the CNN to produce a feature map. The three native features investigated included: carrier frequency (RF), pulse width (PW), and pulse repetition interval (PRI). During pre-processing these features were extracted from the signals and transformed into three different images to be learned by a CNN. This work claims to have properly identified 58 separate emitters with 98.7% accuracy. Secondly, from [7], analyzed two techniques.

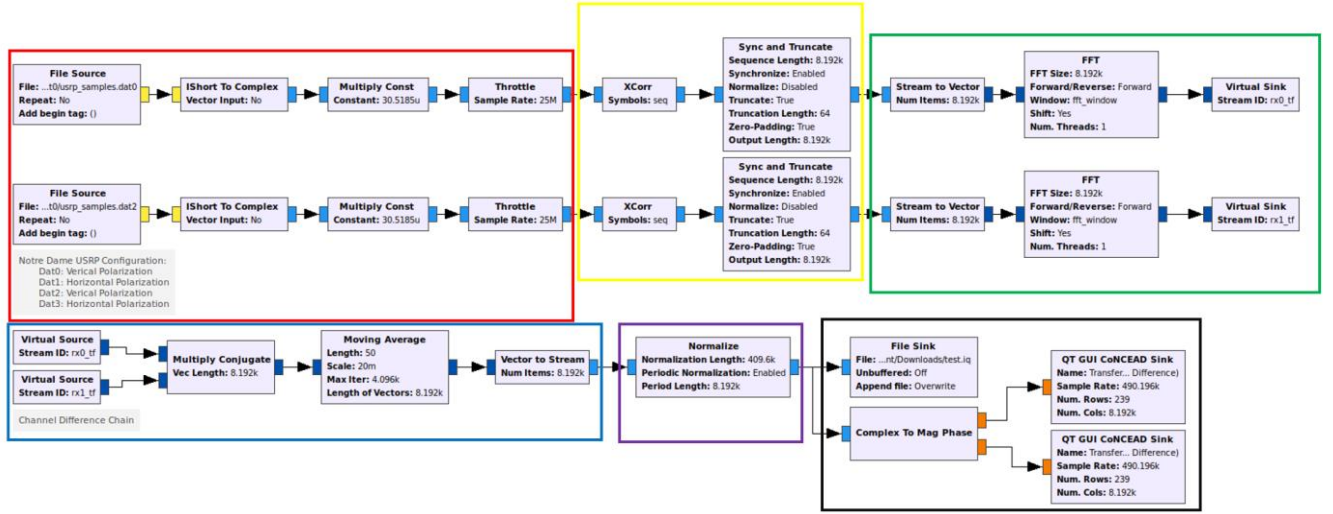

Fig 2: The GNU Radio Flow Graph for the Data Set Extraction for FRV Data Set

One is to use a CNN for the use of identifying emitters using the expert feature of IQ Imbalance. The second approach is to not use any expert features but instead use the inherent capability of the CNN to extract its own features from the raw IQ data for emitter identification. Thirdly, in [10], they present a recurrent neural network (RNN) model that uses the same approach as [8] in terms of features but leverages the RNN's ability to learn certain sequences within the data.

*B. Contributions*

While these related papers and studies have yielded promising results, they have some presented drawbacks. With [8], [9], and [10] they are all using pre-engineered features which are time consuming and require knowledge of certain techniques to obtain the data needed. Moreover, in [8] it requires that you change a 1D representation into a 2D representation which is more computationally complex and would not be the more efficient approach if this application was going to be used on a sensor or small embedded device. While CNNs have traditionally been used for 2D and 3D representations, there have been several advancements over the past five years in the area of 1D CNN applications [10]. This research seeks to capitalize on those advancements.

This paper seeks to use the aforementioned CFs (CIR and TF) for the purpose of EA by way of performing channel identification. According to [12] and [13] the CIR can be used in position estimation of a given emitter target. Furthermore, with the technique of intra-pulse analysis, or the use of both the CFs will yield greater accuracy in the classification and identification of these emitters.

To our knowledge these techniques and approaches have not been used in this way for ES and MIMO system applications. By leveraging CFs of a wide sense stationary (WSS) stochastic processes we can use three unique characteristics to perform EA. 1) CFs can be easily modeled and estimated, 2) CFs are unique to each transmitter, and 3) CFs can be used retro-directively. This investigation produced three questions, hypotheses, and the answers or contributions:

1. Can the unique properties of CFs be used for the association of RF signals?
   H1: In a multi-path channel, the signal travels through distinct paths each with different attenuation, scattering, diffraction, clutter, and other distortions. All of these features are present within the CFs; thus, it gives them unique characteristics for associating the different transmitters to a specific channel and/or direction.
2. Is it possible to build prediction model using a data driven method to extract salient features from CFs that will be able to properly identify channels associated with each emitter?
   H2: A CNN DL based prediction model, based on its ability to create its own feature map from the given data and its ability to exploit spatial dimensions of a wireless channel, can be used for the task of associating RF signals in a MIMO system.
3. Is it possible to use raw CF data that use small pulses of 10000 data points or less to have enough of a difference to perform channel identification?
   H3: The proposed DL prediction model will use the CNNs embedded capability to learn different features one part of the data and identify it is everywhere else within the signal.

## II. METHOD

The proposed DL algorithm for this work is the CNN for raw data identification because of their inherent feature learning abilities and due to their successes in the wireless communications domain in previous work [14] [15]. Though the use of raw data as input to CNN's is a relatively new concept, the success in prior works indicates that CNN's can learn directly from raw signal data. Furthermore, the CNN allows for the input of any size, which gives the algorithm scalability. This in turn allows them to extract features seen in the data, making the algorithm more robust to unknown environments that it will undoubtedly encounter. This research argues for the association of received RF signals based on CFs. Therefore, using the CF matrices, the DL algorithm can be created to use expert features or extract features from the CFs for EA by way of channel identification.

### A. Data Generation & Features

The data for this research was generated using two Ettus X310 USRP's equipped with UBX daughtercards leveraged for transmitting the modulation sequences. Each respective USRP was equipped with two distinct receiving channels and a dual pol antenna. Likewise, the receiver employed a single Ettus X310 USRP equipped with a Twin-Rx daughtercard, thus enabling 4 distinct receive channels. Two spatially separated dual pol antennas mounted to the field research vehicle (FRV) roof rails were used as the receiver aperture in conjunction with the USRP. All 4 receivers are time synchronized and driven with a shared local oscillator. The data collections were recorded to offer multi-path rich environments with line-of-sight between the transmitting and receiving apertures.

The FRV was used as the SDR platform from which all transmitters and receivers in the 4x4 MIMO array were simultaneously controlled. The even numbered receivers used the vertically polarized antenna while the odd numbered leveraged the horizontally polarized antennas. All four receivers were time synchronized and driven with a shared local oscillator. Likewise, four separate transmitters (also referred to as emitters throughout this report) employing two dual pol antennas were used with even numbered transmitters corresponding to vertical antennas and the odd numbered transmitters corresponding to vertical transmitters.

In support of experimentation, a GNU radio flow graph (GNUFG) was created for the data generation, processing of several radar sequences. For the purposes of this research the generated ZC sequences were leveraged with a maximum transmission bandwidth of 12.5 MHz and transmission rate of 25 MHz. Each transmitter transmitted at most two simultaneous ZC sequences, one wide-band (full bandwidth) and either a mid-band (quarter bandwidth) or frequency-hopping narrow-band (one-eight bandwidth). Each transmitter leveraged a unique ZC sequence for each wide-band, mid-band, and narrow-band sequence. These sequences were generated in 10 tests. Test 0 and Test 1 were used as the evaluation data and test 2 through test 9 were used as the training data. While the deep learning (DL) models were trained on test 2-9, the evaluation data was held for testing the models on data that was never seen before for generalization verification. Table 1 summarizes the test sequences.

Table 1. Data Generation Sequences and Combinations

| Transmitter | Sequence | Test 0 | 1 | 2 | 3 | 4 | 5 | 6 | 7 | 8 | 9 |
|---|---|---|---|---|---|---|---|---|---|---|---|
| Tx 0 | Wide-Band | ■ | ■ | ■ |  |  |  | ■ |  |  |  |
|  | Narrow-Band |  |  | ■ | ■ |  |  |  |  |  |  |
| Tx 1 | Wide-Band |  |  |  | ■ | ■ |  |  |  |  |  |
|  | Mid-Band |  |  |  | ■ | ■ |  |  |  |  |  |
| Tx 2 | Wide-Band | ■ | ■ |  |  | ■ | ■ |  |  | ■ |  |
|  | Narrow-Band |  |  |  |  |  | ■ |  |  |  |  |
| Tx 3 | Wide-Band |  |  |  |  |  | ■ | ■ |  |  |  |
|  | Mid-Band |  |  |  |  |  |  | ■ |  |  |  |

### B. CIR and TF Feature Development

For the extraction of the CFs from FRV data set Fig. 2 was used and yielded the data burst shown in Fig. 3 of the comparison of training and evaluation data. Due to the constant amplitude (CA) zero autocorrelation (CAZAC) nature of ZC sequences and negligible cross-correlation properties of sufficiently long and relatively prime-rooted ZC sequences, you can isolate a transmit-receive pair through a simple correlation of these sequences. Furthermore, radar systems have two main reasons that the sequences benefit from CA property. First, a transmitter can operate at peak power if the sequence has a constant peak amplitude. Thus, the system does not have a greater than expected amplitude. Second, amplitude variations during transmission due to additive noise can be theoretically eliminated, which is beneficial when channel classification and estimation is being used. The zero auto-correlation property (ZAC) ensures minimum interference between signals sharing the same channel, which makes signal classification a less difficult task.

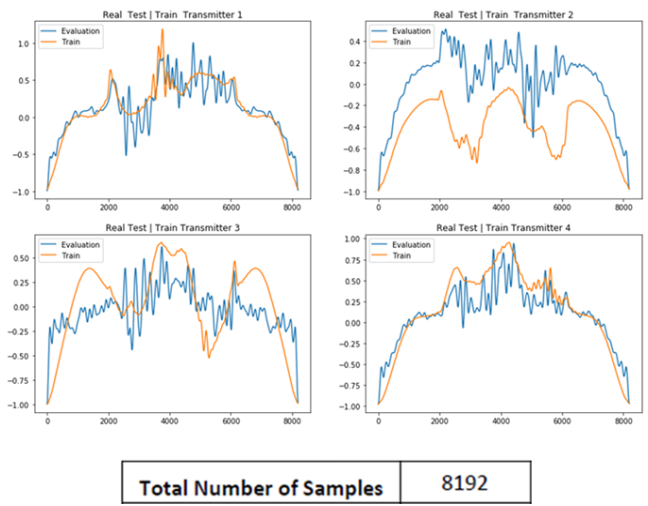

| Total Number of Samples | 8192 |
|---|---|

Fig 3: Example of 8200 Pulse Data Points for Transmitters

Below is an outline of how the TF channel differencing or separation from the GNUFG to obtain the CFs for the dataset is accomplished for the ZC sequence used during the testing at FRV dataset. Each color coordinates to the blocks contained within the corresponding-colored block seen in Fig. 3.

1. Red – File Source and Type Control:
   This is where the file source input is selected. The .DatX files correspond to Receiver/USRP X for each trial. It is important to note that each receiver will record all ZC sequences being transmitted by the active transmitters for each trial.

   The captures were recorded in wire format (interleaved shorts) but are converted to complex floating-point type for processing (hence the change in input/output color from yellow to blue)

2. Yellow - Channel Impulse Response Estimation:
   This performs the correlation against the sequence designated by the parameter selection described for variable B in Figure 12. Thus, the transmit-receive pair's channel estimate or CIR is found via this method. The Sync and Truncate block then synchronizes the output relative to the max correlation (first tap estimation/ first index in a channel response estimate) and truncates/zero pads the output for FFT smoothing.

3. Green - Frequency Domain Transformation:
   Performs FFT to change the output to the frequency domain. Thus, transforming the output from a CIR to the channel TF estimate.

4. Blue - Channel Differencing and Smoothing:
   Uses a correlation between two channel estimate TFs to find the channel difference and a moving average to smooth across time.

5. Purple – Normalization:
   Normalizes the output scaled between 0 and 1 relative to the last number of inputs that correspond to the length attribute.

6. Black - File Writing and Display:
   Writes the output to the designated file in complex floating-point type.

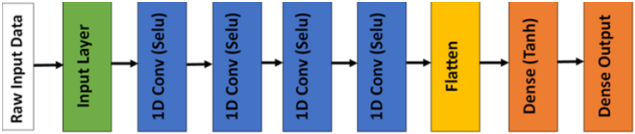

Fig 4: 1D DCNN Model: 4 Conv. Layers, 1 Flatten Layer, Dense Layer with Tanh activation, and Final Dense layer for Classification, SeLU (scaled exponential linear units) Activation, Adam Optimizer

### C. Network Structure

In this research we used two different models to compare the performance to each other and to previous techniques used. The models are 1D Deep Convolutional Neural Network (DCNN) seen in fig 4 and a 4x4 1D DCNN Multi-channel Model (DCNN-MCM) seen in fig. 5. The first model is the traditional single channel DCNN model but without pooling layers. The second model is a DCNN-MCM. The second model was inspired by the hypothesis that every channel for the transmitter-receive pair was unique enough for identification. Given that that was assumed to be the case a model was created to give each channel its own DCNN. In the final layer of the DCNN-MCM model, the global average pooling layer was utilized that is said to help we overfitting the model [16].

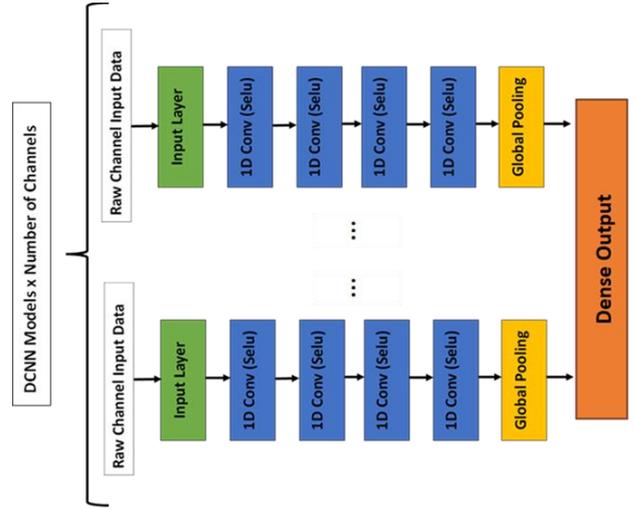

Fig 5: 4 DCNN with 4 Conv. Layers, 1 Global Average, Final Dense Layer for Classification, SeLU Activation, Adam Optimizer

### III. EXPERIMENTAL RESULTS AND DISCUSSION

Both models were trained on 70% of the data, validated with 30% of the data with a batch size of 514, and 8200 samples for each transmitter-received pair. The model's framework was developed and implemented using Anaconda with Keras and Tensorflow. The results of the classification can be seen in fig. 6 and fig. 7 for the DCNN model and the DCNN-MCM respectfully. Both models were trained for an optimal number of epochs based on training and validation comparison. Furthermore, during the experimentation several different activation functions were used however because of its benefits and performance SeLU was used. SeLU, as seen in equation 1, was used after it gave the highest rate of accuracy.

$$selu(x) = \begin{cases} x & , if\ x > 0 \\ \alpha e^x - \alpha, & if\ x \leq 0 \end{cases} \quad (1)$$

SeLU was used based on the characteristics (except for the final layer before the output) is used throughout the algorithm. According to [17], has the following benefits when it comes to neural networks, 1) enables high-level abstract representations, 2) convergence towards zero mean and unit variance even in the presence of noise and perturbations, 3) train deep networks with many layers, 4) employ strong regularization schemes, 5) make learning highly robust, and 6) learn faster.

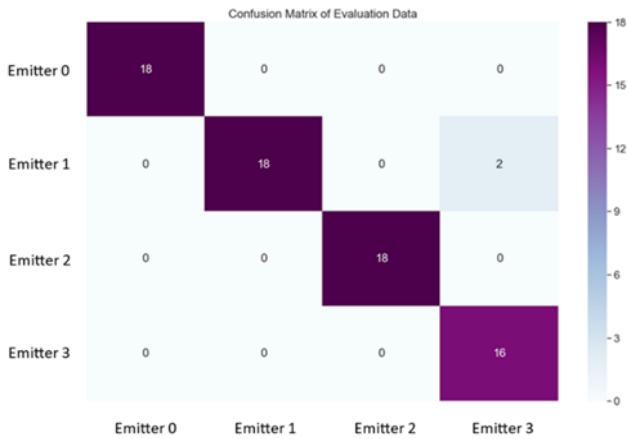

Fig. 6: DCNN Evaluation Data Confusion Matrix

The results provide a first look into the ability for a CNN to accomplish EA using channel identification in a MIMO system using CFs. This technique yielded classification and identification accuracy amongst all emitters of 97.22% and 88.89% accuracy for the DCNN and DCNN-MCM respectively.

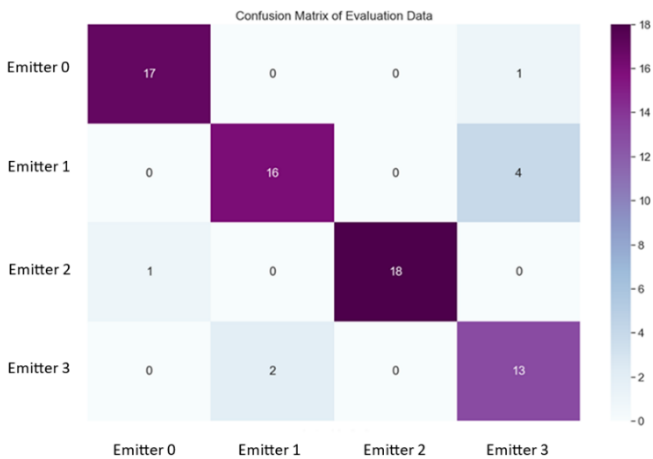

Fig. 7: DCNN-MCM Evaluation Data Confusion Matrix

## IV. CONCLUSION AND FUTURE WORK

This paper presents a novel and efficient method to classify and identify emitter data in a 4x4 MIMO system using the CF's. Adapting the technique of intra-pulse analysis [18] using both the CIR and TF yields promising results for future work and applications. Furthermore, this approach can be applied to many other techniques of classification, identification, and localization. This method successfully classifies and identifies the emitters at 97.22% and 88.89% accuracy for the DCNN and DCNN-MCM respectively.

Moreover, the assumed limitation of this research is the assumption of a static channel. The data was recorded in channels that were static most of the time during the data collection. Given that that is the case, the DL algorithms were not designed to update in real-time. Furthermore, it does not have knowledge of every channel possible through training. Therefore, the future research should look to add real-time updates for future algorithms and the algorithm should be trained on other channel models to have a database of knowledge to leverage in any environment. For example, real-time transfer learning can be used to update trained algorithms with new environmental data. Secondly, this application was based on gaps in military and DoD applications, which would make this research an application that would be beneficial for mobile hand-held applications.